\documentclass[epj]{svjour}
\usepackage{pifont}
\usepackage{amsmath}
\usepackage{txfonts}
\usepackage{graphics}
\usepackage[normalem]{ulem}
\usepackage{xcolor}
\usepackage{amssymb}

\begin{document}
\title{Improved strong-coupling perturbation theory of the symmetric Anderson impurity model}
\author{Kou-Han Ma\inst{1} \and Ning-Hua Tong\inst{1,}
\thanks{\emph{} e-mail: nhtong@ruc.edu.cn}%
}                     
%
%
\institute{Department of Physics, Renmin University of China,
Beijing 100872, P. R. China}
\date{Received: date / Revised version: date}
\abstract{
In a previous work (N. H. Tong, Phys. Rev. B {\bf 92}, 165126 (2015)), an equation-of-motion based series expansion formalism was used to do the second-order strong-coupling expansion for the single-particle Green function of the Anderson impurity model. In this paper, we improve this theory in two aspects. We first use a more accurate scheme to self-consistently calculate the averages that appear in $G_1$. In the resummation process, we use updated coefficients for the continued fraction, guided by the formally exact continued fraction from the Mori-Zwanzig theory. These changes lead to more accurate impurity spin response to the magnetic bias of the bath. Combined with the dynamical mean-field theory, our theory gives improved description for the antiferromagnetism of Hubbard model at half filling.
\PACS{
      {05.30.Jp}   \and
      {05.10.Cc}   \and
      {64.70.Tg}{}
     } 
} 

%

\authorrunning {K.-H. Ma \and  N.-H. Tong}
\titlerunning{Improved strong-coupling perturbation theory of the symmetric Anderson impurity model} \maketitle

\section{Introduction}
The Anderson impurity model (AIM) is one of the best studied quantum many-body models in condensed matter physics. It describes local electron orbitals with Coulomb interaction (impurities) hybridizing with the itinerant non-interacting electron states (bath)~\cite{RefPW:1}.  
AIM has been used to describe metals doped with dilute magnetic impurities~\cite{RefAC:2}, mesoscopic quantum dot systems, molecular conductors~\cite{RefTK:3,RefLI:4,RefYM:4,RefRZ:5,RefRH:5}, and absorption of atoms onto the surface of materials~\cite{RefRB:6,RefDC:7}. In the framework of the dynamical mean-field theory (DMFT), AIM is the effective model for describing the temporal fluctuations of Hubbard model in large spatial dimensions~\cite{RefAG:8}.
Many theoretical and numerical methods are used to solve the AIM, including Bethe-ansatz~\cite{RefPB:6}, perturbation expansion~\cite{RefKY:9,RefSE:10}, Green's function equation of motion~\cite{RefCL:11,RefHG:12}, conserving slave boson theory~\cite{RefJK:13}, noncrossing approximation~\cite{RefNE:13,RefKH:14}, quantum Monte Carlo (QMC)~\cite{RefJE:15,RefMF:16,RefEG:17}, renormalization group~\cite{RefHR:18,RefTA:19,RefRB:20,RefRB:21}, hierarchical equations of motion (EOM)~\cite{RefZH:22}, $etc$. Through decades of efforts, the physics of single impurity AIM has been well understood, though the treatment of multi-orbital AIM is still a challenge~\cite{RefEG:17,RefRB:21}. The single-impurity Anderson model is therefore a priority model for testing the validity of new theoretical methods.

In a previous work~\cite{RefNH:22} (hereafter referred to as I), an EOM based series expansion method for Green's functions (GF) was developed and applied to AIM with a spin bias in the bath. The local GF $G(d_{\sigma}|d_{\sigma}^{\dagger})_{\omega}$ was expanded to the second order of hybridization $V_{k\sigma}$. Resummation was then carried out in the continued fraction (CF) formalism to approximately sum up the series to infinite order, recovering the correct analytical structure of GF. The obtained static averages and local density of states are in quantitative agreement with those from numerical renormalization group (NRG) calculations at intermediate to high temperatures and for small spin bias of the bath. 

However, it is notable that in this theory the impurity spin polarization by the magnetic bias of the bath is significantly underestimated, especially at large $U$, large bias, and low temperature regime. This undermines the value of this theory in the study of quantum dots with ferromagnetic leads and of the magnetic phase of lattice models via DMFT. This deficiency is due to the over simplified approximation, i.e., the atomic approximation, used in I to calculate the higher order static correlation functions. We also found that the CF used for resummation, though recovers the exact form in the atomic limit, does not apply at finite hybridization and arbitrary polarization.
Proper treatment of correlations beyond the $V^{2}_{k\sigma}$ order is therefore vital for quantitative accuracy, even in the strong-coupling regime where a second-order expansion in $V_{k\sigma}$ is supposed to be adequate.

In this paper, we aim to improve the second-order strong-coupling series expansion in the two aspects stated above, by making the following changes to the original theory: 
(i) We do the exact self-consistent calculation for the averages appearing in the first order GF $G_1(d_{\sigma}\vert d_{\sigma}^{\dag})_{\omega}$ (instead of using the atomic approximation for them);
(ii) We use improved CF resummation formalism, containing the exact coefficient $a_1$ from the Mori-Zwanzig theory{~\cite{RefMZ:19} and revised the expression for the coefficient $b_2$;
(iii) All averages are calculated self-consistently from $G(d_{\sigma}|d_{\sigma}^{\dagger})_{\omega}$ alone (instead of being calculated from separate resumed GFs).
With these changes in the theory, we achieve quantitative improvement in the impurity spin polarization under the magnetic bias of bath.

The outline of the paper is as follows. In Sec.2, a brief summary is given for the self-consistent EOM series expansion and the CF resummation. In Sec.3, the improved strong-coupling expansion is carried out for the single-impurity AIM to $V^{2}_{k\sigma}$ order.  In Sec.4, we compare the numerical results from the improved theory, the original theory, and NRG. We also use the improved theory as an impurity solver for DMFT and study the local density of states of the half-filled Hubbard model in the antiferromagnetic phase. The discussion and summary are given in Sec.5.

\section{General formalism}
\label{sec:1}

The general formalism of the EOM-based self-consistent series expansion for GFs was developed in I.
For the sake of completeness, here we give a brief overview.
We consider the following retarded GF of operators $A(t)$ and $B(t')$, 
\begin{equation} \label{eq:1}
G^{r} \left[A(t)\vert B(t') \right] \equiv -i \theta(t-t') \langle [A(t),B(t')]_\pm \rangle .
\end{equation}
Here, $A(t)=e^{iHt}Ae^{-iHt}$.  $[A,B]_{\pm} \equiv AB \pm BA$. $\langle O \rangle$ represents the thermodynamic average of the operator $O$ in the equilibrium state of $H$. Here and below, we use the natural unit $\hbar =1$. The Fourier transform of $G^r[A(t)\vert B(t^{\prime}) ]$ is denoted as
\begin{equation} \label{eq:2}
G(A \vert B)_{\omega + i \eta} \equiv \int_{-\infty}^{+\infty}G^{r} \left[ A(t)\vert B(t') \right] e^{i(\omega + i\eta)(t-t')} d(t-t')
\end{equation}
where $\eta$ is an infinitesimal positive number. On the real frequency axis, the EOM of this double-time GF reads
\begin{eqnarray}\label{eq:3 & 4} 
\omega G(A\vert B)_{\omega} = \langle [A,B]_{\pm} \rangle + G\left( [A,H]\vert B \right)_{\omega},  \label{3}\\
\omega G(A\vert B)_{\omega} = \langle [A,B]_{\pm} \rangle - G\left( A\vert [B,H] \right)_{\omega}.  \label{4}
\end{eqnarray}
Below, we only use the Fermion-type GF with the anti-commutator $[A,B]_{+} = \{A, B\}$ in the EOM.

\subsection {Self-consistent series expansion}

We decompose the given Hamiltonian $H$ as $H = H_0 + \lambda H_1$. $H_0$ is the exactly solvable part and $H_1$ is regarded as the perturbation. $\lambda$ is the formal expansion parameter and will be set as unity after the calculation. 
To develop an EOM-based self-consistent series expansion for GF, we formally write
\begin{eqnarray}  \label{eq:5}
 G(A\vert B)_{\omega} &=& G_0(A\vert B)_{\omega} + G_1(A\vert B)_{\omega}+ \cdots + G_n(A\vert B)_{\omega} 
 \nonumber\\&&+ \Gamma_{n}(A\vert B)_{\omega}. 
\end{eqnarray}
Here, $G_{i}(A|B)_{\omega} \propto \lambda^{i}$ ($i=0,1,...,n$) is the $i$-th order GF. $\Gamma_{n}(A\vert B)_{\omega}$ is the residue of this expansion at order $n$ .

From the left-side EOM Eq.(3), the zeroth-order GF is defined by the EOM
\begin{equation}   \label{eq:6}
\omega G_0(A\vert B)_{\omega} = \langle \{A, B \} \rangle + G_0([A,H_0]\vert B)_{\omega}.
\end{equation}
Here, $\langle \{ A, B \} \rangle$ is the full average to be calculated self-consistently after the GF is formally obtained.
By putting Eq.(5) into Eq.(3) and comparing the two sides of equation order by order, we obtain the EOM for
the $i$-th order GF $G_{i}(A|B)_{\omega}$ ( $i {\geqslant} 1$) as
\begin{equation}   \label{eq:7}
\omega G_i(A\vert B)_{\omega} = G_{i-1}([A,H_1]\vert B)_{\omega} + G_i([A,H_0]\vert B)_{\omega}.
\end{equation}
Each order of GF can be solved exactly if the commutator series $[A, H_0]$, $[[A, H_0], H_0]$, $...$ closes automatically.
The $n$th-order residue $\Gamma_{n}(A\vert B)_{\omega}$ satisfies
\begin{equation}  \label{eq:8}
\omega \Gamma_{n}(A\vert B)_{\omega} = G_{n}([A,H_1]\vert B)_{\omega} + \Gamma_{n}([A,H]\vert B)_{\omega},
\end{equation}
which cannot be solved exactly in general.
The series expansion could be deduced also from the right-side EOM Eq.(4), which gives different $G_i(A\vert B)_{\omega}$'s. Hereafter, we will use the series from the left-side EOM, Eqs.(6)-(8). This theory is named self-consistent series expansion.

\subsection{CF resummation}
A physically acceptable GF should be Lehmann representable, $i.e.$, it consists of real simple poles. 
Suppose we have obtained the self-consistent series expansion of $G(A\vert B)_\omega$ up to the $m$-th order, 
\begin{equation} \label{eq:9}
G(A\vert B)_{\omega} \approx G_{0}(A\vert B)_{\omega} + G_{1}(A\vert B)_{\omega} + \cdots + G_{m}(A\vert B)_{\omega}.
\end{equation}
Using the CF resummation~\cite{RefSP:23}, we can obtain the GF $G_{CF}(A|A^{\dagger})_{\omega}$ from Eq.(9) (taking $B=A^{\dagger}$) 
\begin{equation}   \label{eq:10}
G_{CF}(A|A^{\dagger})_{\omega}  = \cfrac{a_0}{\omega + b_1
        - \cfrac{a_1}{\omega + b_2 - ...   } }. 
\end{equation}
It was proven that for $a_l$, $b_{l+1}$ real and $a_l \geqslant 0$ ($l=0,1,...$), Eq.(\ref{eq:10}) is Lehmann representable~\cite{RefJG:24}. In order to determine $a_l$ and $b_l$, we can expand $G(A\vert A^{\dagger})_{\omega}$ and $G_{CF}(A\vert A^{\dagger})_{\omega}$ into Taylor series of $1/\omega$ and require that for every $ n \in [1,+\infty]$, the $\omega^{-n}$ terms of $G(A\vert B)_{\omega}$ and $G_{CF}(A\vert B)_{\omega}$ equal on the level of $\lambda^m$. Here,  $m$ is the order of perturbation. In this paper, we take $\lambda = V_k$ and $m=2$.
To calculate the averages (all real in this work) appearing in the expansion, we use the fluctuation-dissipation theorem,
\begin{equation} \label{eq:11}
\langle BA \rangle = -\frac{1}{\pi}\int_{-\infty}^{\infty}\text{Im} G(A\vert B)_{\varpi+i\eta}\frac{1}{e^{\beta\varpi}+1}d\varpi.
\end{equation}

\section{Strong-coupling expansion for Anderson impurity model}

 We consider the single-impurity Anderson model
\begin{equation} \label{eq:12}
H_{AIM} = \sum_{k\sigma}\epsilon_{k\sigma}c_{k\sigma}^\dag c_{k\sigma} + U n_\uparrow n_\downarrow + \epsilon_d \sum_{\sigma}n_\sigma + \sum_{k\sigma} V_{k\sigma}(c_{k\sigma}^\dag d_{\sigma} + h.c.).
\end{equation}
Here, $c_{k\sigma}$/$d_{\sigma}$ is the bath/impurity electron annihilation operator. We split $H_{AIM}=H_0 + H_1$ as
\begin{equation} \label{eq:13}
H_0 = \sum_{k\sigma}\epsilon_{k\sigma}c_{k\sigma}^\dag c_{k\sigma} + U n_\uparrow n_\downarrow + \epsilon_d \sum_{\sigma}n_{\sigma} , 
\end{equation}
and
\begin{equation} \label{eq:14}
H_1 = \sum_{k\sigma} V_{k\sigma}(c_{k\sigma}^\dag d_{\sigma} + h.c.).
\end{equation}
$H_0$ is exactly solvable due to the decoupling of impurity from bath. $H_1$ is treated as a perturbation.
For the hybridization function $\Delta_{\sigma}(\omega)\equiv\sum_{k}V_{k\sigma}^2 \delta(\omega -\epsilon_{k\sigma})$, we use a Lorentzian form
\begin{equation}   \label{eq:15}
\Delta_{\sigma}(\omega)=\frac{\Gamma\omega_{c}^2}{(\omega+\sigma\delta\omega)^2+\omega_{c}^2}.
\end{equation}
Here, $\Gamma$ is the strength of the hybridization. $\delta\omega$ is the spin bias on the bath electrons, which is used to describe the magnetic electrode in quantum dot systems or the magnetic phase of a lattice Hamiltonian in DMFT. $\sigma=1$ for spin up and $-1$ for spin down. $\omega_{c}$ is set as the energy unit.
As in paper I, here we only focus on the particle-hole symmetric case. Generalization to the particle-hole asymmetric case is straightforward.
For the above hybridization function that fulfils $\Delta_{\sigma}(-\omega)=\Delta_{\bar{\sigma}}(\omega)$, the particle-hole symmetric point is located at $\epsilon_d = - U/2$. We also define two related intermediate functions
\begin{eqnarray}   \label{eq:16}
\Gamma_{\sigma}(\omega)&=& \int_{-\infty}^{\infty}\frac{\Delta_{\sigma}(\epsilon)}{\omega-\epsilon} d\epsilon, \nonumber\\
\Lambda_{\sigma}(\omega)&=& \int_{-\infty}^{\infty}\frac{\Delta_{\sigma}(\epsilon)}{\omega-\epsilon}\frac{1}{e^{\beta\epsilon}+1} d\epsilon.
\end{eqnarray}
 They fulfil the relations $\Gamma_{\sigma}(-\omega)=-\Gamma_{\bar{\sigma}}(\omega)$ and $\Lambda_{\sigma}(-\omega)=\Lambda_{\bar{\sigma}}(\omega)-\Gamma_{\bar{\sigma}}(\omega)$.

\subsection{Second order $V_{k\sigma}$ expansion of GF}

In I, a self-consistent expansion of the impurity GF was developed up to $V_{k\sigma}^{2}$ order in the standard basis operator (SBO) formalism. In this paper, these expressions are kept intact and we only modify the schemes for the subsequent resummation and self-consistent calculation. For completeness, in this part we summarize the notations and results of I.

The standard basis operators (SBOs)~\cite{RefSB:25,RefHL:26} used in I are the excitation operators of the local Hamiltonian $h_0 = U n_\uparrow n_\downarrow + \epsilon_d \sum_{\sigma}n_\sigma$. Denoting $|\mu \rangle$ ($\mu = 1 \sim 4$) as the eigenstates of $h_{0}$, i.e., $h_0 | \mu \rangle = E_{\mu} \vert\mu\rangle$, we have $\vert 1\rangle=d_{\uparrow}^{\dag}\vert 0\rangle$, $\vert 2\rangle=d_{\downarrow}^{\dag}\vert 0\rangle$, $\vert 3\rangle=\vert 0\rangle$, and $\vert 4\rangle=d_{\uparrow}^{\dag}d_{\downarrow}^{\dag}\vert 0\rangle$. The corresponding eigen energies are $E_1=E_2=\epsilon_d$, $E_3=0$, and $E_4=U+2\epsilon_d$. The SBOs $\{A_{\alpha\beta}\}$ are defined as $A_{\alpha\beta}\equiv\vert\alpha\rangle\langle\beta\vert$. Obviously, we have  
$\sum_{\alpha}A_{\alpha\alpha}=1$ and $A_{\alpha\beta}A_{\mu\nu}=\delta_{\beta\mu}A_{\alpha\nu}$. The impurity annihilation operator is expressed as $d_{\sigma}=\sum_{\mu \nu}f_{\mu\nu}^{\sigma} A_{\mu\nu}^{\sigma}$. $f^{\sigma}_{\mu\nu} =0$ except for 
$f_{31}^{\uparrow}=f_{24}^{\uparrow}=1$ and $f_{32}^{\downarrow}=-f_{14}^{\downarrow}=1$. 
Below we use $A^{\sigma}_{\mu\nu}$ to denote the SBO which net annihilates or creates an electron with spin $\sigma$. We use $A_{\mu\nu}$ without superscript to denote the SBO that changes the number of electrons by an even number. Therefore, $A^{\sigma}_{\mu\nu}$ is Grassman odd and $A_{\mu\nu}$ is Grassman even. We have $\{ A_{\mu\nu}^{\sigma}, c_{k\sigma} \}=0$ and $\left[ A_{\mu\nu}, c_{k \sigma} \right] = 0$. 

With these definitions, we have
\begin{eqnarray}  \label{eq:17}
 && \{A_{\alpha\beta}^{\sigma},d_{\sigma'}^{\dag}\}=\sum_{\mu\nu}M_{\alpha\beta,\mu\nu}^{\sigma\sigma'}A_{\mu\nu},  \nonumber \\
&& \{A_{\alpha\beta}^{\sigma},d_{\sigma'}\}=\sum_{\mu\nu}N_{\alpha\beta,\mu\nu}^{\sigma\sigma'}A_{\mu\nu},
\end{eqnarray}
and
\begin{eqnarray}   \label{eq:18}
&&  M_{\alpha\beta,\mu\nu}^{\sigma\sigma'}=\delta_{\mu\alpha}
f_{\nu\beta}^{\sigma'\ast}+\delta_{\nu\beta}
f_{\alpha\mu}^{\sigma'\ast},   \nonumber \\
&& N_{\alpha\beta,\mu\nu}^{\sigma\sigma'}=\delta_{\mu\alpha}
f_{\beta\nu}^{\sigma'}+\delta_{\nu\beta}
f_{\mu\alpha}^{\sigma'}.
\end{eqnarray}
The commutators between SBOs and $H_0$ and $H_1$ are respectively
\begin{equation}   \label{eq:19}
\left[ A_{\alpha\beta}^{\sigma},H_0 \right]=(E_{\beta}-E_{\alpha})A_{\alpha\beta}^{\sigma},
\end{equation}
and 
\begin{equation}   \label{eq:20}
\left[ A_{\alpha\beta}^{\sigma},H_1 \right]=\sum_{k\sigma'}\sum_{\mu\nu}V_{k\sigma'} \left( M_{\alpha\beta,\mu\nu}^{\sigma\sigma'}A_{\mu\nu}c_{k\sigma'}-N_{\alpha\beta,\mu\nu}^{\sigma\sigma'}c_{k\sigma'}^{\dag}A_{\mu\nu} \right).
\end{equation}
The impurity GF $G(d_{\sigma}\vert d_{\sigma}^{\dag})_{\omega}$ can be decomposed into
$G(d_{\sigma}\vert d_{\sigma}^{\dag})_{\omega}=\sum_{\alpha\beta\gamma\delta}f_{\alpha\beta}^{\sigma}f_{\gamma\delta}
^{\sigma\ast}G(A_{\alpha\beta}^{\sigma}\vert A_{\gamma\delta}
^{\sigma\dag})_{\omega}$. The self-consistent strong-coupling expansion 
\begin{eqnarray}   \label{eq:21}
&& G(A_{\alpha\beta}^{\sigma}\vert A_{\gamma\delta}^{\sigma\dag})_{\omega} \approx G_{0}(A_{\alpha\beta}^{\sigma}\vert A_{\gamma\delta}^{\sigma\dag})_{\omega} + G_{1}(A_{\alpha\beta}^{\sigma}\vert A_{\gamma\delta}^{\sigma\dag})_{\omega} + G_{2}(A_{\alpha\beta}^{\sigma}\vert A_{\gamma\delta}^{\sigma\dag})_{\omega}  \nonumber \\
&&
\end{eqnarray}
was obtained in I. We will not repeat the expressions but refer the readers to I for details: Eq.(84) for $G_{0}(A_{\alpha\beta}^{\sigma}\vert A_{\gamma\delta}^{\sigma\dag})_{\omega}$, Eq.(88) for $G_{1}(A_{\alpha\beta}^{\sigma}\vert A_{\gamma\delta}^{\sigma\dag})_{\omega}$, and Eqs.(89)-(91) for $G_{2}(A_{\alpha\beta}^{\sigma}\vert A_{\gamma\delta}^{\sigma\dag})_{\omega}$.  

\subsection{Improved self-consistent calculation}

These GF components in Eq.(21) were expressed in terms of the unknown averages such as $\langle A_{\delta\beta} \rangle$ (in $G_{0}$), $\langle A_{\gamma\mu}^{\sigma\dag}c_{k\sigma'}\rangle$ (in $G_{1}$), and $\langle A_{\lambda\gamma}c_{p\sigma''} c_{k\sigma'}\rangle$, $\langle A_{\lambda\gamma}c_{p\sigma''}^{\dag} c_{k\sigma'}\rangle$ (in $G_{2}$).
In I, they were calculated from the following minimum approximations that ensures Eq.(\ref{eq:21}) being exact at the order $V_{k\sigma}^2$.

$\langle A_{\delta\beta}\rangle$ in $G_0$ was calculated from the resummed GF $G_{CF}(A_{\alpha\beta}^{\sigma}\vert A_{\gamma\delta}^{\sigma\dag})_{\omega}$. 
The averages of the type $\langle Ac \rangle$ in $G_1$ were calculated from the corresponding GFs at $V_{k\sigma}$ order. For instance,  $\langle A_{\gamma\mu}^{\sigma\dag}c_{k\sigma'} \rangle$ was calculated via the fluctuation-dissipation theorem from the approximate GF 
\begin{eqnarray}   \label{eq:22}
G(c_{k\sigma'}\vert A_{\gamma\mu}^{\sigma\dag})_{\omega}&\approx & G_0(c_{k\sigma'}\vert A_{\gamma\mu}^{\sigma\dag})_{\omega}
+G_1(c_{k\sigma'}\vert A_{\gamma\mu}^{\sigma\dag})_{\omega}\nonumber\\
&=&\frac{V_{k\sigma'}}{\omega - \epsilon_{k\sigma'}} 
G_0(d_{\sigma^{\prime}} \vert A_{\gamma\mu}^{\sigma\dag})_{\omega}.
\end{eqnarray}
Other averages in $G_1$ can be obtained by symmetry transformations from it.

The averages of the type $\langle Acc \rangle$ or $\langle Ac^{\dag}c \rangle$ in $G_2$ describe the spin exchange and pair hopping between impurity and bath, being important for the description of Kondo effect. In I, they were calculated by a truncation valid at the order $V_k^0$,
\begin{eqnarray}   \label{eq:23}
\langle A_{\lambda\gamma}c_{p\sigma''}c_{k\sigma'}\rangle\approx
\langle A_{\lambda\gamma}\rangle \langle c_{p\sigma''}c_{k\sigma'}\rangle_{0} = 0 , \nonumber\\
\langle c_{p\sigma''}^{\dag}c_{k\sigma'}A_{\lambda\gamma}\rangle
 \approx \langle c_{p\sigma''}^{\dag}c_{k\sigma'}\rangle_{0} \langle A_{\lambda\gamma}\rangle.
\end{eqnarray}
$\langle O \rangle_0$ means the zero-th order average.
Eqs.(\ref{eq:22}) and (\ref{eq:23}) guarantees that the truncated series of $G(A_{\alpha\beta}^{\sigma}\vert A_{\gamma\delta}^{\sigma\dag})
_{\omega}$ is exact at $V_{k\sigma}^2$ order.

As analyzed in the introduction, merely keeping the exact $V_{k\sigma}^2$ order is not sufficient for certain purpose even in the strong-coupling limit, for which the accuracy of higher order terms are vital.
In this section, we go one step forward than the previous minimum approximation. Instead of using the approximation Eq.(\ref{eq:22}), we calculate $\langle A_{\gamma\mu}^{\sigma\dag}c_{k\sigma'} \rangle$ in $G_1$ from the following exact relation
\begin{equation}   \label{eq:24}
G(c_{k\sigma'}\vert A_{\gamma\mu}^{\sigma\dag})_{\omega}= \frac{V_{k\sigma'}}{\omega - \epsilon_{k\sigma'}}G(d_{\sigma'}
\vert A_{\gamma\mu}^{\sigma\dag})_{\omega},
\end{equation}
with $G(d_{\sigma'} \vert A_{\gamma\mu}^{\sigma\dag})_{\omega}$ to be calculated self-consistently from the CF-resummed impurity GFs. Compared with Eq.(\ref{eq:22}), the present scheme treats higher order hybridization effect more accurately.

With the new self-consistent scheme Eq.(\ref{eq:24}), we obtain an updated expression for $G_1(A_{\alpha\beta}^{\sigma}\vert A_{\gamma\delta}^{\sigma\dag})_{\omega}$ as 
\begin{equation}    \label{eq:25}
G_1(A_{\alpha\beta}^{\sigma}\vert A_{\gamma\delta}^{\sigma
\dag})_{\omega}=\frac{K_{\alpha\beta,\gamma\delta}^{\sigma}(\omega)}{\omega+E_{\alpha}-E_{\beta}},
\end{equation}
with
\begin{eqnarray}   \label{eq:26}
K_{\alpha\beta,\gamma\delta}^{\sigma}(\omega) &= &
\sum_{\sigma'}\sum_{\mu\nu}M_{\alpha\beta,\mu\nu}^{\sigma\sigma'}[\delta_{\mu\gamma}\Phi_{\delta\nu}^{\sigma'}(\omega+E_{\mu}-E_{\nu})-\nonumber\\&&\delta_{\nu\delta}\Phi_{\mu\gamma}^{\sigma'}(\omega+E_{\mu}-E_{\nu})]\nonumber\\&-
&\sum_{\sigma'}\sum_{\mu\nu}N_{\alpha\beta,\mu\nu}^{\sigma\sigma'}[\delta_{\mu\gamma}\Phi_{\nu\delta}^{\sigma'}(-\omega-E_{\mu}+E_{\nu})-\nonumber\\&&\delta_{\nu\delta}\Phi_{\gamma\mu}^{\sigma'}(-\omega-E_{\mu}+E_{\nu})].
\end{eqnarray}
The function $\Phi_{\mu \gamma}^{\sigma}(\omega)$ is given by
\begin{eqnarray}   \label{eq:27}
\Phi_{\mu\gamma}^{\sigma}(\omega) &\equiv &  \sum_{k}\frac{V_{k
\sigma}\langle A_{\mu\gamma}^{\sigma}c_{k\sigma}\rangle}{\omega-\epsilon_{k\sigma}}  \nonumber\\
&=& \int_{-\infty}^{\infty} d\epsilon \frac{\Delta_{\sigma}(\epsilon)}{\omega-\epsilon} \Big\langle \frac{G(d_{\sigma}\vert A_{\gamma\mu}^{\sigma\dag})_{\varpi}}{\varpi-\epsilon} \Big\rangle.
\end{eqnarray}
Here, for a given function $g(\varpi)$, we define $\langle g(\varpi)\rangle \equiv -(1/\pi) \int_{-\infty}^{\infty}{\text Im} g(\varpi + i \eta)/ (e^{\beta \varpi} + 1) d\varpi$. 
If we replace $G(d_{\sigma} \vert A_{\gamma\mu}^{\sigma\dag})_{\varpi}$ in Eq.(\ref{eq:27}) with the zero-th order quantity $G_0(d_{\sigma} \vert A_{\gamma\mu}^{\sigma\dag})_{\varpi}$, Eq.(\ref{eq:25}) reduces to the old result, Eq.(97) of I. Therefore, the new scheme Eq.(\ref{eq:24}) modifies only the $V_{k\sigma}^{n \geqslant 3}$ order contributions in $G_1(A_{\alpha\beta}^{\sigma}\vert A_{\gamma\delta}^{\sigma\dag})_{\omega}$ but keeps the $V_{k\sigma}^{n \leqslant 2}$ order intact. 
Note that Eq.(\ref{eq:25}) has an explicit pole at $E_{\beta}-E_{\alpha}$, while the corresponding old one, Eq.(97) of I, has two explicit poles. This suggests that although $K_{\alpha\beta,\gamma\delta}^{\sigma}(\omega)$ in Eq.(\ref{eq:26}) contains real simple poles apparently, an additional pole emerges in the expansion to $V_{k\sigma}^{2}$ order. This could make a second-order pole in $K_{\alpha\beta,\gamma\delta}^{\sigma}(\omega)$ in the strong coupling limit. Due to this observation, when we do CF resummation, $K_{\alpha\beta,\gamma\delta}^{\sigma}(\omega)$ should not be regarded as a constant (see below).

For the averages in $G_2(A_{\alpha\beta}^{\sigma}\vert A_{\gamma\delta}^{\sigma\dag})_{\omega}$, in this paper we still use the old scheme Eq.(\ref{eq:23}). The resulting expression for $G_2(A_{\alpha\beta}^{\sigma}\vert A_{\gamma\delta}^{\sigma\dag})_{\omega}$ is not changed and the expression was given by Eq.(98) of I. 
We will discuss how to improve Eq.(\ref{eq:23}) in Sec.5.

\subsection{Improved CF resummation}

Besides the self-consistent calculation, there is room for improvement also in the CF resummation process.
At the particle-hole symmetric point, the second-order truncated series of the single-particle local GF is simplified into (from Eq.(\ref{eq:21}))
\begin{eqnarray}  
G(d_{\sigma}\vert d_{\sigma}^{\dag})_{\omega}&\approx& G_0(d_{\sigma}\vert d_{\sigma}^{\dag})_{\omega}+G_1(d_{\sigma}\vert d_{\sigma}^{\dag})_{\omega}
+G_2(d_{\sigma}\vert d_{\sigma}^{\dag})_{\omega}  \label{eq:28}
 \\
&=&\frac{W_1^{\sigma}}{\omega + U/2}+\frac{W_2^{\sigma}}{\omega - U/2}+\frac{W_3^{\sigma}(\omega)}{(\omega + U/2)^2}\nonumber\\ &+&\frac{W_4^{\sigma}(\omega)}{(\omega - U/2)^2}+\frac{W_5^{\sigma}(\omega)}{(\omega + U/2)(\omega - U/2)}.  \label{eq:29}
\end{eqnarray}
The expressions for $W_1^{\sigma} \sim W_5^{\sigma}(\omega)$ are summarized in Appendix A. Due to the improved self-consistent calculation scheme Eq.(24), $W_{3 \sim 5}^{\sigma}(\omega)$ are different from those in I. The following exact relations still hold, $W_{1}^{\sigma} + W_{2}^{\sigma}=1$, $W_{1}^{\sigma} - W_{2}^{\sigma}=1 -2 \langle n_{\bar{\sigma}} \rangle$, and $W_3^{\sigma}(\omega)+ W_4^{\sigma}(\omega)+W_5^{\sigma}(\omega)=\Gamma_{\sigma}(\omega)$. They give out certain exact properties of GF such as sum rule and the exactness in the non-interacting limit.

We used the following two level extended CF to do the resummation for Eq.(\ref{eq:29}), 
\begin{equation}    \label{eq:30}
G_{CF}(d_{\sigma}\vert d_{\sigma}^{\dag})_{\omega}= \cfrac {a_0}{\omega + b_1(\omega) - \cfrac {a_1}{\omega + b_2(\omega)} }.
\end{equation}
Here, $a_0$ and $a_1$ are real and positive $\omega$-independent constants. $G_{CF}(d_{\sigma}\vert d_{\sigma}^{\dag})_{\omega}$ is Lehmann representable if $b_1(\omega)$ and $b_2(\omega)$ contain only real simple poles. Eq.(\ref{eq:30}) is an extension of Eq.(\ref{eq:10})  to $\omega$-dependent $b_i$'s, which can produce a continuous spectral function from a finite level CF. In I, we expanded Eqs.(\ref{eq:29}) and (\ref{eq:30}) into Taylor series of $1/\omega$ and equated the corresponding coefficients to produce
\begin{eqnarray}  \label{eq:31}
&& a_{0} = 1, \nonumber \\
&& b_{1}(\omega) = \frac{U}{2}\left( W_1^{\sigma}-W_2^{\sigma} \right) - \Gamma_{\sigma}(\omega),  \nonumber \\
&& a_{1}+b_1(\omega)^2 = \left(\frac{U}{2} \right)^2 -U \left[ W_3^{\sigma}(\omega)-W_4^{\sigma}(\omega) \right],  \nonumber\\
&& 2 a_1 b_1(\omega) + a_1 b_2(\omega)+ b_1(\omega)^3  \nonumber \\
&& = \left( \frac{U}{2}\right)^3 \left[ W_1^{\sigma}-W_2^{\sigma} \right]- 3 \left( \frac{U}{2} \right)^2 \Gamma_{\sigma}(\omega)  + 2 \left( \frac{U}{2} \right)^2 W_5^{\sigma}(\omega). \nonumber \\
&&
\end{eqnarray}
In the expansion, $W_{3 \sim 5}^{\sigma}(\omega)$ and $b_{1 \sim 2}(\omega)$ are regarded as constants and the exact relations among $W_{i}^{\sigma}$'s were used. Since Eq.(29) contains second-order pole at most and hence a two level extended CF is sufficient. In I, the coefficients $a_{i}$'s and $b_{i}(\omega)$'s were solved from the above equations at the $V_{k\sigma}^{2}$ order. In particular, in solving the third equation of Eq.(31) for $a_1$, $b_1(\omega)^2$ and $W_{3}^{\sigma}(\omega)-W_{4}^{\sigma}(\omega)$ were discarded since they are at the order $V_{k\sigma}^4$ (Note $\langle n_{\bar{\sigma}}\rangle = 1/2 + \mathcal{O}(V_{k\sigma}^{2})$). This produced
\begin{eqnarray}   \label{eq:32}
&& a_0 = 1, \,\,\,\,\,\,\,\,\, a_1 = \left(\frac{U}{2}\right)^2, \nonumber \\
&& b_1(\omega)=\frac{U}{2}(W_1^{\sigma}-W_2^{\sigma})-\Gamma_{\sigma}(\omega),  \nonumber \\
&& b_2(\omega) = -\frac{U}{2}(W_1^{\sigma}-W_2^{\sigma})-\Gamma_{\sigma}(\omega) + 2 W_5^{\sigma}(\omega).
\end{eqnarray}
The $b_1(\omega)$ and $b_2(\omega)$ obtained above contain real simple poles only and satisfy the Lehmann representability requirement.

The above resummation scheme used in I has some degrees of freedom to modify. First, the forms of $W_{1}^{\sigma} \sim W_{5}^{\sigma}(\omega)$ cannot be uniquely determined by the expression of Eq.(\ref{eq:29}). Second, the terms of order $V_{k\sigma}^{n \geqslant 3}$ could be kept in the final $a_{i}$ ($i=0,1$) or $b_{i}(\omega)$ ($i=1,2$) to improve the accuracy.
In this work, we first find a suitable expression for $W_{i}^{\sigma}$'s which are now different from those of I. 
Considering that in Eq.(25), $K_{\alpha\beta,\gamma\delta}^{\sigma}(\omega)$ contains the second-order pole at $\omega =  E_{\delta}-E_{\gamma}$ in the strong coupling limit, we multiply $\left( \omega  + E_{\gamma} - E_{\delta} \right)$ simultaneously on the nominator and the denominator of Eq.(25). This leads to the modified expression for $W_i^{\sigma}$'s in Appendix A, which in any cases only have real simple poles. In mapping GF to CF, these $W_i^{\sigma}$'s are regarded as constants.

Second, in solving the third equation of Eq.(\ref{eq:31}) for $a_1$, instead of neglecting $b_1(\omega)^2$ completely, we now keep the static part of it $U/2(W_1^{\sigma} - W_{2}^{\sigma}) = U/2(1 - 2 \langle n_{\bar{\sigma}} \rangle)$. This is motivated by the observation that though each coefficient in Eq.(\ref{eq:32}) is exact at the $V_{k\sigma}^2$ order, the response of $\langle n_{\sigma} \rangle$ to $\delta \omega$ is unsatisfactory, especially at large $U$~\cite{RefNH:22}. The exact local self-energy in the atomic limit  $\Sigma_{\sigma}(\omega) = \epsilon_d + U \langle n_{\bar{\sigma}}\rangle + U^2 \langle n_{\bar{\sigma}} \rangle (1-\langle n_{\bar{\sigma}} \rangle)/\left[ \omega - \epsilon_d - U(1-\langle n_{\bar{\sigma}} \rangle ) \right]$ reminds us that to accurately describe the polarization of impurity spin by the bias of bath, one needs to use the full term $a_1 = U^2 \langle n_{\bar{\sigma}} \rangle (1-\langle n_{\bar{\sigma}} \rangle )$ instead of the $V_{k\sigma}^2$ order one $a_1 = (U/2)^2$. Preserving the full term $U/2(W_1^{\sigma} - W_{2}^{\sigma})$ in the static part of $b_{1}(\omega)$ recovers the atomic self-energy for arbitrary impurity polarization.

Using the above modifications, we obtain the improved CF coefficients as
\begin{eqnarray}   \label{eq:33}
&& a_{0}=1, \,\,\,\,\,\,\,\, a_1=U^2\langle n_{\bar{\sigma}}\rangle(1-\langle n_{\bar{\sigma}}\rangle),  \nonumber\\
&& b_1(\omega)=\frac{U}{2}(W_1^{\sigma}-W_2^{\sigma})-\Gamma_{\sigma}(\omega),  \nonumber \\
&& b_2(\omega)= -\frac{U}{2}(W_1^{\sigma}-W_2^{\sigma}) -\Gamma_{\sigma}(\omega) +\frac{1}{2\langle n_{\bar{\sigma}}\rangle(1-\langle n_{\bar{\sigma}}\rangle)} W_5^{\sigma}(\omega).   \nonumber \\
&&
\end{eqnarray}
Eq.(\ref{eq:33}) satisfies the Lehmann representability and recovers Eq.(\ref{eq:32}) in the unbiased limit ($\delta \omega =0$). In the biased case, Eq.(\ref{eq:33}) is expected to describe the impurity polarization more accurately.

The improved CF resummation scheme can be justified by the exact extended CF formalism of the local GF of AIM. Using the Mori-Zwanzig operator projection method~\cite{RefMZ:19}, we obtain the formally exact expression
\begin{eqnarray}   \label{eq:34}
&& G(d_{\sigma}\vert d_{\sigma}^{\dag})_{\omega}  \nonumber \\
&& = \cfrac{1}{\omega - \Omega_0 - \Gamma_{\sigma}(\omega)
-\cfrac{U^2\langle n_{\bar{\sigma}}\rangle(1-\langle n_{\bar{\sigma}}\rangle)}{\omega -\Omega_1 
+\frac{\langle n_{\bar{\sigma}}\rangle}{1-
\langle n_{\bar{\sigma}}\rangle}\Gamma_{\sigma}(\omega)
+iK_{2c}(-i\omega) } }.   \nonumber \\
&& 
\end{eqnarray}
The frequencies in the above equation are given by $\Omega_{0}=\epsilon_d + U\langle n_{\bar{\sigma}}\rangle$ and $\Omega_{1}=\epsilon_d + U\left( 1-\langle n_{\bar{\sigma}}\rangle \right) + \tilde{\beta}_{\sigma}$, with 
\begin{equation} \label{eq:35}
\tilde{\beta}_{\sigma} = \frac{\sum_{k}V_{k\bar{\sigma}}\langle c_{k\bar{\sigma}}^{\dag}d_{\bar{\sigma}}\left(2n_{\sigma}-1\right)\rangle }{ \langle n_{\bar{\sigma}}\rangle \left(1-\langle n_{\bar{\sigma}}\rangle \right) }.
\end{equation}
The memory function $iK_{2c}(-i\omega)$ is a GF of higher order operators, whose EOM cannot be solved exactly. The derivation of Eq.(34) is summarized in Appendix B.
Comparing Eq.(33) to Eq.(34), we see that our improved CF resummation gives the exact coefficients $a_0$, $a_1$, and $b_1(\omega)$. For $b_2(\omega)$, the constant term is also exact, given $W_{5}^{\sigma}(\omega = \infty) = 2 \sum_{k}V_{k\bar{\sigma}} \langle c_{k\bar{\sigma}}^{\dag}d_{\bar{\sigma}}\left(1-2n_{\sigma} \right)\rangle$. The $\omega$-dependence of $b_{2}(\omega)$ is approximate. Eq.(34) is also recovered at various levels by the recently developed projective truncation approximations with different operator bases~\cite{Fan1,Fan2}.

The improved second-order strong-coupling expansion formulas involve the unknown averages $\langle A_{\alpha\alpha}\rangle$ and the functions $\Phi_{\alpha\beta}^{\sigma}(\omega)$. To calculate them self-consistently, we first write $\langle A_{11}\rangle=\langle n_{\uparrow}\rangle-\langle n_{\uparrow}n_{\downarrow}\rangle$, $\langle A_{22}\rangle=\langle n_{\downarrow}\rangle-\langle n_{\uparrow}n_{\downarrow}\rangle$, $\langle A_{33}\rangle=\langle A_{44}\rangle=(1-\langle A_{11}\rangle-\langle A_{22})/2$. To calculate $\langle n_{\uparrow}n_{\downarrow}\rangle$ and $\Phi_{\alpha\beta}^{\sigma}(\omega)$ , we use 
\begin{equation}   \label{eq:36}
G(d_{\sigma}\vert n_{\bar{\sigma}}d_{\sigma}^{\dag})_{\omega}=
\frac{\omega-\Gamma_{\sigma}(\omega)-\epsilon_{d}}{U}G(d_{\sigma}\vert d_{\sigma}^{\dag})_{\omega}-1/U .
\end{equation}
Consequently, all averages are calculated from the CF-resummed single-particle GF $G_{CF}(d_{\sigma}\vert d_{\sigma}^{\dag})_{\omega}$. This is different from I where the CF-resummed GFs $G_{CF}(A_{31}^{\uparrow}\vert A_{31}^{\uparrow\dag})_{\omega}$, $G_{CF}(A_{24}^{\uparrow}\vert A_{24}^{\uparrow\dag})_{\omega}$, $etc$ are required.

\section{ Numerical results}

In this section, we present numerical results for the improved strong-coupling expansion (denoted as ISC below) and compare them with those from the original strong-coupling expansion (denoted as SC below) and NRG. We use the full density matrix NRG algorithm~\cite{RefAW:27} supplemented with the self-energy trick for the local density of states (LDOS)~\cite{RefRB:20} to provide a reference for comparison. We use logarithmic discretization parameter $\Lambda=2.0$ and keep $M_{s}=[356,380]$ states in the NRG calculation~\cite{RefZH}.

\begin{figure}
\vspace{-2.0cm}
\begin{center}
\resizebox{0.8\textwidth}{!}{
\includegraphics{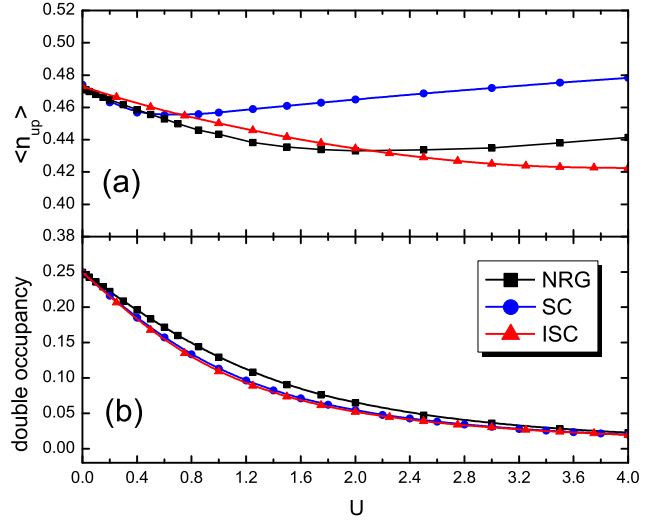}}
\vspace{-1.0cm}
\end{center}
\caption{$\langle n_{\uparrow}\rangle$  and  $\langle n_{\uparrow}n_{\downarrow}\rangle$ as functions of U for a spin-polarized bath. Model parameters are $\Gamma=0.1, \delta\omega=0.2$, and $T=0.1$.} \label{fig:1}
\end{figure}

Fig.\ref{fig:1} presents impurity electron occupancies $\langle n_{\uparrow}\rangle$ and the double occupancy $\langle n_{\uparrow}n_{\downarrow}\rangle$ as functions of $U$ for a spin-polarized bath at $T=0.1$. In Fig.\ref{fig:1}(a), it is seen that $\langle n_{\uparrow}\rangle$ calculated by the two methods are in good agreement with NRG in the small $U$ regime. For SC, the quantitative agreement is maintained only upto $U=0.5$. The quantity is less accurate for ISC in the small $U$ regime but the better qualitative agreement sustains to larger interaction $U\sim 2.0$. In larger $U$ regime, both results deviate from NRG significantly, with ISC overestimating and SC underestimating the polarization, respectively. In Fig.\ref{fig:1}(b), the double occupancy obtained by the two methods are all in good agreement with NRG, except for slight deviations at medium $U$ value.
\begin{figure}
\vspace{-2.0cm}
\begin{center}
\resizebox{0.8\textwidth}{!}{
\includegraphics{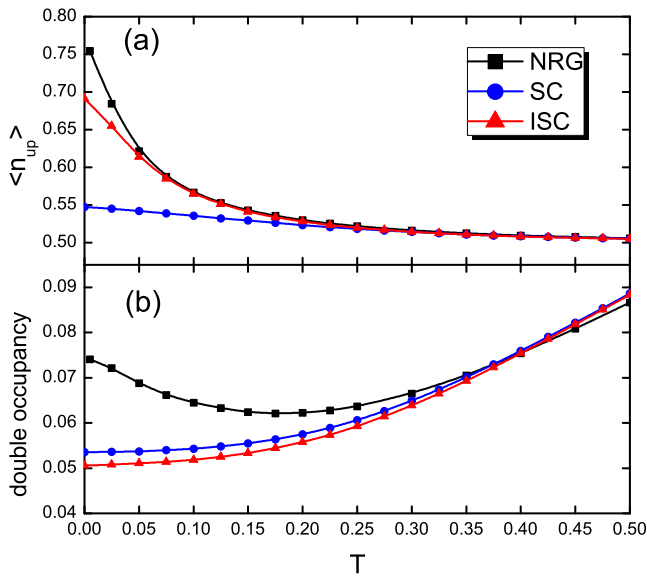}}
\vspace{-1.0cm}
\end{center}
\caption{$\langle n_{\uparrow}\rangle$ and $\langle n_{\uparrow}n_{\downarrow}\rangle$ as functions of T for a spin-polarized bath. Model parameters are $\Gamma=0.1, \delta\omega=-0.2$, and $U=2.0$.}\label{fig:2}
\end{figure}

Fig.\ref{fig:2} presents $\langle n_{\uparrow}\rangle$ and $\langle n_{\uparrow}n_{\downarrow}\rangle$ as functions of temperature for a negative bias $\delta\omega = -0.2$ and intermediate interaction $U=2.0$. In Fig.\ref{fig:2}(a), the value of $\langle n_{\uparrow} \rangle$ from ISC is almost same as that of NRG except for $T < 0.1$, showing dramatical improvement over SC. In Fig.\ref{fig:2}(b), $\langle n_{\uparrow} n_{\downarrow}\rangle$ from ISC has no improvement but decreases slightly from the SC result at low temperatures. Neither ISC nor SC produces the upturn of $\langle n_{\uparrow} n_{\downarrow}\rangle$ in the NRG data for $T < 0.2$ which is associated with the establishment of Kondo screening and Fermi liquid state. Such deviation is expected because neither SC nor ISC can accurately describe the Kondo effect at low temperatures due to the crude truncation approximation Eq.(\ref{eq:23}) for the averages of the type $\langle Acc \rangle$ and $\langle Ac^{\dag}c \rangle$ in $G_2(A_{\alpha\beta}^{\sigma}\vert A_{\gamma\delta}^{\sigma\dagger})_{\omega}$. 

\begin{figure}
\vspace{-2.0cm}
\begin{center}
\resizebox{0.8\textwidth}{!}{
\includegraphics{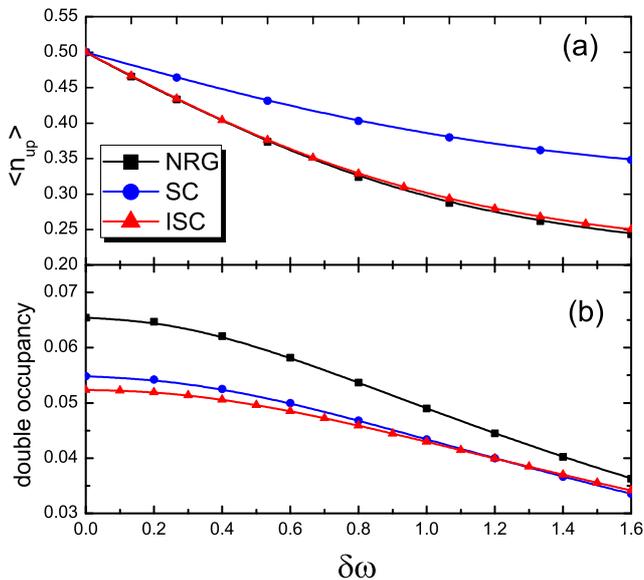}}
\vspace{-1.0cm}
\end{center}
\caption{$\langle n_{\uparrow}\rangle$ and $\langle n_{\uparrow}n_{\downarrow}\rangle$ as functions of $\delta\omega$. Model parameters are $\Gamma=0.1, T=0.1$, and $U=2.0$.}   \label{fig:3}
\end{figure}
Fig.\ref{fig:3} shows $\langle n_{\uparrow}\rangle$ and $\langle n_{\uparrow}n_{\downarrow}\rangle$ as functions of $\delta\omega$ for an intermediate $U=2.0$. In Fig.\ref{fig:3}(a), ISC produces a curve of $\langle n_{\uparrow}\rangle(\delta \omega)$ almost identical to NRG upto the large bias regime, significantly surpassing SC. Similar to Fig.2(b), the accuracy of the double occupancy data from ISC deteriorates slightly with respect to SC, both being much smaller than that of NRG. This shows that ISC, with improved evaluation of the averages of type $\langle Ac \rangle$ and with the exact CF coefficients $a_1$, only improves the response of the impurity spin to the bias of bath but gains little in the hybridization-induced correlation effect, i.e., the Kondo physics. As stated above, such effect is encoded in the averages of the type $\langle Acc \rangle$ and $\langle Ac^{\dag}c \rangle$ in $G_2(A_{\alpha\beta}^{\sigma}\vert A_{\gamma\delta}^{\sigma\dagger})_{\omega}$.
\begin{figure}
\vspace{-2.0cm}
\begin{center}
\resizebox{0.8\textwidth}{!}{
\includegraphics{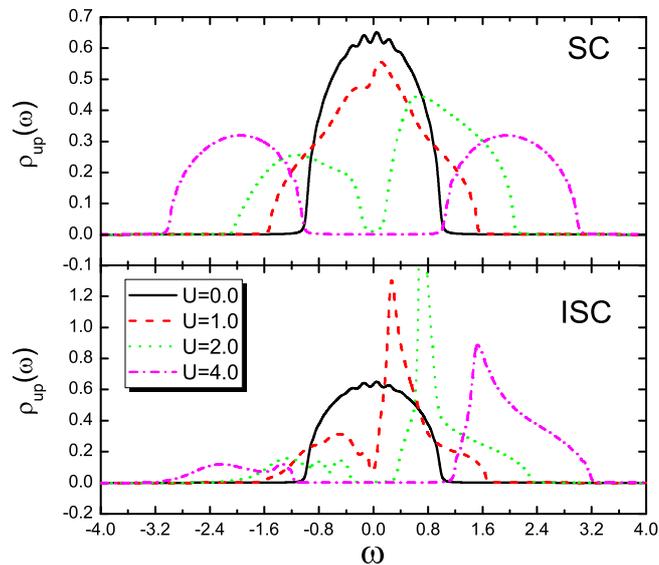}}
\vspace{-1.0cm}
\end{center}
\caption{Local density of states of Hubbard model on the half-filled Bethe lattice: $\rho_{\uparrow}(\omega)$ for various $U$ values, obtained from (a) SC, and (b) ISC at $T=0.01$. We use the antiferromagnetic initialization with $\delta\omega=0.1$. From top to bottom at $\omega=0$: $U = 0.0$, $1.0$, $2.0$, and $4.0$.}    \label{fig:4}
\end{figure}
To further demonstrate the improvement of ISC over SC, we use them as impurity solvers to study the antiferromagnetism of the half-filled Hubbard model within DMFT. The description of antiferromagnetism in DMFT relies crucially on the accurate description of the impurity density of states under the bias of bath. As is well known, many strong-coupling based approximations such as the Hubbard-I approximation and the alloy analogy approximation cannot give out the antiferromagnetic phase of the Hubbard model at half filling~\cite{RefGebhard}. ISC is expected to do better than SC in this respect due to the improved description of the impurity spin response to the bias of bath. The Hubbard model Hamiltonian reads  
\begin{equation}    \label{eq:37}
H=-t\sum_{ \langle i,j \rangle} \sum_{\sigma} \left( c_{i\sigma}^{\dagger}c_{j\sigma}+h.c. \right) +U \sum_{i} n_{i\uparrow}n_{i\downarrow}.
\end{equation}
DMFT is exact for this model on the Bethe lattice with coordination number $z=\infty$, which has a semicircular free density of states~\cite{RefAG:8}
\begin{equation}    \label{eq:38}
 D(\epsilon)=\frac{2}{\pi W^2}\sqrt{W^2-\epsilon^2}.
\end{equation}
$W$ is the half-bandwidth which is set as the energy unit. We used the standard DMFT self-consistent equations for the Bethe lattice with $z=\infty$ in the antiferromagnetic phase~\cite{RefAG:8}. 

Fig.\ref{fig:4} shows the LDOS of the half-filled Hubbard model at a low temperature $T=0.01$, obtained from SC (Fig.4(a)) and ISC (Fig.4(b)), respectively. 
At $U=0$, the LDOS's from SC and ISC are both exact, corresponding to a paramagnetic metal phase. With gradual increase of $U$, distinctions appear in two aspects. First, the LDOS from SC opens an energy gap only at intermediate $U$ while that from ISC opens a gap almost immediately as $U > 0$. This contrast is seen clearly in the $U=1.0$ curves: the SC curve is clearly a metal while the ISC curve has a dip at $\omega=0$, being more close to a true insulator. The ISC results agree better with the fact that the ground state of Hubbard model on a bipartite lattice is a Slater insulator in the weak coupling regime, due to the Fermi surface nesting~\cite{RefAG:8,RefTP:27}. We checked that the small finite $\rho_{\uparrow}(\omega=0)$ in the ISC curve of $U=1.0$ is not due to finite temperature, but due to the ISC approximation. 

In the second aspect, for intermediate to large $U$ regimes, SC only gives out very weak antiferromagnetism as shown by the weakly asymmetric shape of the LDOS. The antiferromagnetism disappears for $U \geqslant 4.0$ in Fig.4(a). In contrast, in Fig.4(b), ISC gives much more robust antiferromagnetism with a strongly asymmetric LDOS and a sharp peak on the shoulder of the upper sub-band. The antiferromagnetism persists to the large $U$ limit. This shows that ISC correctly produces the antiferromagnetic ground state of this system in the large $U$ regime. The N\'{e}el temperature $T_N$ obtained from ISC varies with $U$ in the same qualitative trend as the QMC data. However, both the ground state order parameter and the magnitude of N\'{e}el temperature are overestimated by the ISC solver, compared to the QMC results~\cite{RefQMC:28}. This observation is consistent with the poor description of the Kondo screening by ISC since the unscreened local moments order easier and generate stronger magnetism.

\section{Discussions and Summary}

Besides the improved self-consistent calculation of the averages of the type $\langle Ac \rangle$ in $G_1(A_{\alpha\beta}^{\sigma}\vert A_{\gamma\delta}^{\sigma\dagger})_{\omega}$, we  also tried to improve the truncation approximation Eq.(\ref{eq:23}) for the averages of the types $\langle Acc \rangle$ and $\langle Ac^{\dagger}c \rangle$ in $G_2(A_{\alpha\beta}^{\sigma}\vert A_{\gamma\delta}^{\sigma\dagger})_{\omega}$. For an example, for $\langle A_{\lambda\gamma}c_{p\sigma''}c_{k\sigma'}\rangle$, we  use $\langle A_{\lambda\gamma}c_{p\sigma''}c_{k\sigma'}\rangle= \Big\langle G(c_{k\sigma'}\vert A_{\lambda\gamma}
c_{p\sigma''})_{\omega} \Big\rangle$ and expand the latter to first-order of $V_{k\sigma}$, generating
\begin{eqnarray}    \label{eq:39}
G(c_{k\sigma'}\vert A_{\lambda\gamma}c_{p\sigma''})_{\omega}&\approx &
\frac{V_{k\sigma'}}{\omega -\epsilon_{k\sigma'}}\sum_{\alpha
\beta}f_{\alpha\beta}^{\sigma'}G_0(A_{\alpha\beta}^{\sigma'}\vert A_{\lambda\gamma}c_{p\sigma''})_{\omega}  \nonumber \\
&=&\frac{V_{k\sigma'}}{\omega -\epsilon_{k\sigma'}}\sum_{\alpha
\beta}f_{\alpha\beta}^{\sigma'}\frac{\delta_{\lambda\beta}\langle A_{\alpha\gamma}^{\sigma'}c_{p\sigma''}\rangle - \delta_{\alpha\gamma}\langle A_{\lambda\beta}^{\sigma'}c_{p\sigma''}\rangle}{\omega + E_{\alpha}-E_{\beta}}. \nonumber \\
&&
\end{eqnarray}
The averages of the type $\langle Ac \rangle$ in the second equation are calculated self-consistently by scheme Eq.(\ref{eq:24}).
Compared to the truncation approximation Eq.(\ref{eq:23}) which is exact at $V_{k\sigma}^{0}$, the above approximation is exact at $V_{k\sigma}^{2}$ and partially takes into account the spin exchange between impurity and the bath electrons.
Replacing Eq.(\ref{eq:23}) with this approximation, we obtain the improved $G_2(A_{\alpha\beta}^{\sigma}\vert A_{\gamma\delta}^{\sigma\dagger})_{\omega}$
which modifies the expression of functions $F_{\alpha\beta,\gamma\delta}^{\sigma}(\omega)$ in $W_{3 \sim 5}^{\sigma}(\omega)$. 

However, we find that this improvement, together with the improvement in calculating $\langle Ac \rangle$ (Eq.(\ref{eq:24})) and in the CF resummation (Eq.(\ref{eq:33})), does not produce uniformly improved results over ISC. This may be attributed to the fact that the Kondo singlet are formed by degenerate many-body states and is hence singular at $V_{k \sigma}=0$. Any expansion from the atomic limit will break the spin exchange process. To handle the Kondo screening accurately, a truly non-perturbative calculation of the averages of the types $\langle Acc \rangle$ and $\langle Ac^{\dagger}c \rangle$ are necessary. We note that the decoupling approximation of the Lacroix type~\cite{RefCL:11}, such as $\langle AcAc\rangle \approx \langle Ac \rangle \langle Ac\rangle$ was proved to be able to describe the Kondo effect. Exploration in this direction will be carried out in the forthcoming work.

One of the motivations of the present study is to develop an impurity solver for the multi-orbital AIM, which is of more practical interest in the community of DMFT. An accurate and fast multi-orbital impurity solver is valuable for DMFT study of strongly correlated $d$ or $f$ electron materials. 
Most of our formula used in this paper can be generalized to multi-orbital AIM, due to the generality of the SBO formalism.
However, the CF resummation will become more complicated for multi-orbital AIM because the zero-th order GF contains larger number of poles instead of two in the single-orbital case. Consequently, a multi-level extended CF must be used to do the resummation. This makes the analytical derivation more difficult, if not impossible at all. Therefore, a computer-aided CF resummation or the matrix generalization of the present strong-coupling expansion theory are required for this purpose. 
Note that our formalism, though demonstrated in this paper only for the particle-hole symmetric case, can be generalized to particle-hole asymmetric case without problem.

Besides the EOM-based series expansion of GF used in this paper, the Mori-Zwanzig CF formalism~\cite{RefMZ:19,RefLee} is also a useful framework to carry out systematic strong-coupling expansion, as was done for Hubbard model on the Bethe lattice in Ref.\cite{RefHong}. 
In the present work, the GF is first expanded into a series of a small parameter and an extended CF is then constructed from the truncated series. Ref.\cite{RefHong} used an inverse procedure, i.e., one first write down the exact formal expression for the CF via the recursive relation, and then the coefficients of this CF are expanded into a series and truncated to a given order. It thus avoids the difficulty of mapping GF to CF in our approach. The exact formal expression for $G(d_{\sigma}|d_{\sigma}^{\dagger})_{\omega}$ Eq.(\ref{eq:34}) obtained from the Mori-Zwanzig formalism suggests that this approach has certain advantages and may provide an alternative to the present approach. In both approaches, the self-consistent calculation of averages that encode the important ground state correlation is unavoidable. Quantitative comparison of the two approaches are under study.

In summary, we improve the second-order EOM-based strong-coupling perturbation theory for the symmetric AIM, which was first developed in I. The improvement consists of two aspects. First, the averages of the type $\langle Ac \rangle$ in $G_1(A_{\alpha\beta}^{\sigma}\vert  A_{\gamma\delta}^{\sigma\dag})_{\omega}$ is calculated from an exact relation instead of a $V_{k\sigma}^{1}$ order approximation. Second, contributions from $V_{k\sigma}^{n \geqslant 3}$ are retained in the CF coefficients $a_1$ which now assumes the exact form. The averages are calculated self-consistently through the CF-resummed GF $G_{CF}(d_{\sigma}\vert d_{\sigma}^{\dag})_{\omega}$. Using NRG results as reference, we show that these modifications significantly improve the response of the impurity spin to the bath bias. Combined with DMFT, improved description is obtained for the antiferromagnetic phase of the half-filled Hubbard model on Bethe lattice. We also present an extended CF formalism for the GF of AIM, which may be the starting point of an alternative strong-coupling expansion theory.

\section{Acknowledgements}
This work is supported by 973 Program of China (2012CB921704), NSFC grant (11374362), Fundamental Research Funds for the Central Universities, and the Research Funds of Renmin University of China 15XNLQ03.

\appendix{}

\section{Appendix: expressions for $W_1^{\sigma} \sim W_5^{\sigma}(\omega)$}

In Appendix A, we give the explicit expression for $W_1^{\sigma} \sim W_5^{\sigma}(\omega)$ in Eq.(29). To obtain them, we follow Eqs.(84), (88)-(91) of I and insert the self-consistent calculation schemes Eq.(23) and (24) into them. We also multiplied $\left( \omega + E_{\gamma} - E_{\delta}\right)$ to $K^{\sigma}_{\alpha \beta, \gamma \delta}(\omega)$ to take care of the Lehmann representability of $W_{i}^{\sigma}(\omega)$ ($i=3,4,5$). At particle-hole symmetric point, we obtain for spin up,
\begin{eqnarray}   \label{eq:40}
&&\hspace{-0.5em}W_1^{\uparrow}=\langle A_{11}\rangle + \langle A_{33}\rangle,   \nonumber \\
&&\hspace{-0.5em}W_2^{\uparrow}=\langle A_{22}\rangle + \langle A_{44}\rangle,    \nonumber\\
&&\hspace{-0.5em}W_3^{\uparrow}(\omega)=F_{31,31}^{\uparrow}(\omega)+\left(\omega + \frac{U}{2}\right) K_{31,31}^{\uparrow}(\omega), \nonumber\\
&&\hspace{-0.5em}W_4^{\uparrow}(\omega)=F_{24,24}^{\uparrow}(\omega)+\left( \omega - \frac{U}{2} \right) K_{24,24}^{\uparrow}(\omega),  \nonumber \\
&&\hspace{-0.5em}W_5^{\uparrow}(\omega)=F_{31,24}^{\uparrow}(\omega)+F_{24,31}^{\uparrow}(\omega)+ \left(\omega + \frac{U}{2} \right) K_{24,31}^{\uparrow}(\omega)  \nonumber\\
&&\hspace{3.5em}  + \left(\omega - \frac{U}{2} \right) K_{31,24}^{\uparrow}(\omega). 
\end{eqnarray}
For spin down, we obtain
\begin{eqnarray}    \label{eq:41}
&&\hspace{-0.5em}W_1^{\downarrow}=\langle A_{22}\rangle + \langle A_{33}\rangle,  \nonumber \\
&&\hspace{-0.5em} W_2^{\downarrow}=\langle A_{11}\rangle + \langle A_{44}\rangle,  \nonumber\\
&&\hspace{-0.5em} W_3^{\downarrow}(\omega)=F_{32,32}^{\downarrow}(\omega)+ \left(\omega + \frac{U}{2} \right) K_{32,32}^{\downarrow}(\omega),  \nonumber\\
&&\hspace{-0.5em} W_4^{\downarrow}(\omega)=F_{14,14}^{\downarrow}(\omega)+ \left( \omega - \frac{U}{2} \right) K_{14,14}^{\downarrow}(\omega),   \nonumber \\
&&\hspace{-0.5em}W_5^{\downarrow}(\omega)=-F_{32,14}^{\downarrow}(\omega)-F_{14,32}^{\downarrow}(\omega)- \left(\omega + \frac{U}{2} \right) K_{14,32}^{\downarrow}(\omega),   \nonumber\\
&&\hspace{3.5em}- \left(\omega - \frac{U}{2} \right) K_{32,14}^{\downarrow}(\omega).  
\end{eqnarray}
The functions $K^{\sigma}_{\alpha\beta, \gamma\delta}(\omega)$ in the above equations are given by
\begin{eqnarray}   \label{eq:42}
&&\hspace{-0.5em} K_{31,31}^{\uparrow}(\omega)\hspace{-0.1em}=-K_{24,31}^{\uparrow}(\omega) \hspace{-0.1em} = -\Phi_{23}^{\downarrow}(\omega)\hspace{-0.1em}+\hspace{-0.1em}\Phi_{41}^{\downarrow}(-\omega), \nonumber \\
&&\hspace{-0.5em} K_{24,24}^{\uparrow}(\omega)\hspace{-0.1em}=-K_{31,24}^{\uparrow}(\omega)\hspace{-0.1em}=\Phi_{23}^{\downarrow}(-\omega)\hspace{-0.1em}-\hspace{-0.1em}\Phi_{41}^{\downarrow}(\omega),   \nonumber\\
&&\hspace{-0.5em} K_{32,32}^{\downarrow}(\omega)\hspace{-0.1em}=K_{14,32}^{\downarrow}(\omega)\hspace{-0.1em}=-\Phi_{13}^{\uparrow}(\omega)\hspace{-0.1em}-\hspace{-0.1em}\Phi_{42}^{\uparrow}(-\omega),   \nonumber \\
&&\hspace{-0.5em} K_{14,14}^{\downarrow}(\omega)\hspace{-0.1em}= K_{32,14}^{\downarrow}(\omega)\hspace{-0.1em} = \Phi_{13}^{\uparrow}(-\omega)\hspace{-0.1em}+\hspace{-0.1em}\Phi_{42}^{\uparrow}(\omega).
\end{eqnarray}
The functions $F^{\sigma}_{\alpha\beta, \gamma\delta}(\omega)$ are given by
\begin{eqnarray}    \label{eq:43}
&&\hspace{-0.5em}F_{31,31}^{\uparrow}(\omega)=I_{13}[\Gamma_{\uparrow}(\omega)+\Lambda_{\downarrow}(\omega)-\Lambda_{\downarrow}(-\omega)], \nonumber\\
&&\hspace{-0.5em}F_{24,24}^{\uparrow}(\omega)=I_{24}[\Gamma_{\uparrow}(\omega)+\Gamma_{\downarrow}(\omega)-\Gamma_{\downarrow}(-\omega)-\Lambda_{\downarrow}(\omega)+\Lambda_{\downarrow}(-\omega)], \nonumber\\
&&\hspace{-0.5em}F_{32,32}^{\downarrow}(\omega)=I_{23}[\Gamma_{\downarrow}(\omega)+\Lambda_{\uparrow}(\omega)-\Lambda_{\uparrow}(-\omega)], \nonumber\\
&&\hspace{-0.5em}F_{14,14}^{\downarrow}(\omega)=I_{14}[\Gamma_{\downarrow}(\omega)+\Gamma_{\uparrow}(\omega)-\Gamma_{\uparrow}(-\omega)-\Lambda_{\uparrow}(\omega)+\Lambda_{\uparrow}(-\omega)], \nonumber \\
&&
\end{eqnarray}
and
\begin{eqnarray}    \label{eq:44}
&&\hspace{-0.5em}F_{24,31}^{\uparrow}(\omega)=I_{13}\Gamma_{\uparrow}(\omega)-F_{31,31}^{\uparrow}(\omega), \nonumber \\
&&\hspace{-0.5em}F_{31,24}^{\uparrow}(\omega)=I_{24}\Gamma_{\uparrow}(\omega)-F_{24,24}^{\uparrow}(\omega),  \nonumber\\
&&\hspace{-0.5em}F_{14,32}^{\downarrow}(\omega)=F_{32,32}^{\downarrow}(\omega)-I_{23}\Gamma_{\downarrow}(\omega), \nonumber \\
&&\hspace{-0.5em}F_{32,14}^{\downarrow}(\omega)=F_{14,14}^{\downarrow}(\omega)-I_{14}\Gamma_{\downarrow}(\omega).
\end{eqnarray}
Here, $I_{\alpha \beta} \equiv \langle A_{\alpha\alpha}\rangle + \langle A_{\beta\beta}\rangle$. The particle-hole symmetry gives the relations $\Phi_{13}^{\uparrow}(\omega) = \Phi_{41}^{\downarrow}(-\omega)$ and $\Phi_{42}^{\uparrow}(\omega)=- \Phi_{23}^{\downarrow}(-\omega)$.

\section{Appendix: extended CF formula for $G(d_{\sigma}|d_{\sigma}^{\dagger})_{\omega}$ }

In Appendix B, we derive the formally exact extended CF formula Eq.(34) for $G(d_{\sigma}|d_{\sigma}^{\dagger})_{\omega}$ of AIM, using the Mori-Zwanzig formalism.

We consider the Heisenberg equation of motion for the operator $A_0(t) = e^{iHt} A_0 e^{-iHt}$ in the Heisenberg picture,
\begin{eqnarray}     \label{eq:45}
i\frac{d}{d t}A_0(t)=[A_0(t),H].
\end{eqnarray}
Using the Liouville operator $LO\equiv[H,O]$, we get
\begin{eqnarray}     \label{eq:46}
A_0(t)=e^{iLt}A_0.
\end{eqnarray}
Here and below, $O(t)$, $O(z)$, and $O$ are used to denote the Heisenberg operator, its Laplace transformation, and $O(t=0)$, respectively.
We introduce the inner product $(A\vert B)$ of the operator $A$ and $B$ and define the projection operator $P_1$ as
\begin{eqnarray}    \label{eq:47}
P_1 O \equiv \frac{(A_0 \vert O)}{(A_0\vert A_0)}A_0.
\end{eqnarray}
$P_1$ is a superoperator that projects any operator into the one-dimensional subspace of $A_0$. The projection operator of the 
orthogonal subspace is $Q_1 = 1-P_1$. We have the relations $P_1^2=P_1$, $Q_1^2=Q_1$, and $P_1Q_1=Q_1P_1=0$. In this work, we use the inner product $(A\vert B)=\langle \{A^{\dagger},B\}\rangle$, which guarantees $P_1=P_1^{\dagger}$, $Q_1=Q_1^{\dagger}$, and $L=L^{\dagger}$.
The generalized Langevin equation for $A_0(t)$ reads
\begin{equation}    \label{eq:48}
\frac{d}{dt} A_0(t)
= -i\Omega_0 A_0(t) + iA_1(t) - \int_{0}^{t} d \tau A_0(t-\tau) K_1(\tau).
\end{equation}
It is an alternative formulation of the Heisenberg equation of motion Eq.(\ref{eq:45}). $\Omega_0 = (A_0\vert [A_0,H])/(A_0\vert A_0)$ is the frequency and the random force is given by $A_1(t) = e^{iQ_1 Lt}A_1$ with $A_1 = Q_1 L A_0$. $A_1(t)$ is orthogonal to $A_0$. The memory function reads $K_1(t)= (A_1\vert A_1(t))/(A_0\vert A_0)$.
The derivation of Eq.(\ref{eq:48}) can be found in Ref.\cite{RefMZ:19} where the use of the following Dyson identity is made
\begin{eqnarray}    \label{eq:49}
e^{iLt}=e^{iQ_1Lt} +  i\int_{0}^{t} d \tau e^{iL(t-\tau)}P_1Le^{iQ_1L\tau}.
\end{eqnarray}
Applying the Laplace transformation $f(z) = \int_0^{\infty} f(t) exp(-zt) dt$ to Eq.(46) and Eq.(48), we obtain respectively the Heisenberg equation of motion and the generalized Langevin equation on the complex variable $z$ axis
\begin{eqnarray}  \label{eq:50}
&& z A_0(z) = A_0 + i \left[LA_0 \right](z),  \nonumber \\
&& A_0(z)= \frac{A_0 + i A_1(z)}{z + i\Omega_0 + K_1(z)}.
\end{eqnarray}

Projecting the second equation of Eq.(\ref{eq:50}) to $A_0$, we obtain 
\begin{equation}  \label{eq:51}
 K_0(z)\equiv (A_0 \vert A_0(z)) = \frac{(A_0\vert A_0)}{z+i\Omega_0 + K_1(z)}.
\end{equation}
The Fermion-type Matsubara Green's function $G(A_0\vert A_0^{\dagger})_{i\omega_n}$ is obtained  from $K_0(z)$ via 
$G(A_0\vert A_0^{\dagger})_{i\omega_n} = -i K_0(z = \omega_n)$.

To calculate $K_1(z)$, one needs to obtain $A_1(t)$. It satisfies the equation of motion and the (equivalent) generalized Langevin equation,
\begin{eqnarray}
   \frac{d A_1(t)}{ dt} &=& i Q_1 L A_1(t)   \\  \label{eq:52}
   &=& -i \Omega_1 A_1(t) + i A_2(t) - \int_0^{t} A_1(t-\tau) K_2(\tau).   \label{eq:53}  
\end{eqnarray}
In Eq.(53), $\Omega_1 = (A_1\vert [A_1,H]))/(A_1\vert A_1)$ and $A_2(t) = e^{iQ_2 Q_1 Lt}A_2$ with $A_2 = Q_2 Q_1 L A_1$. The memory function reads $K_2(t)= (A_2\vert A_2(t))/(A_1\vert A_1)$. The projection operator $Q_2 = 1 - P_2$ and $P_2 O = \left[(A_1| O)/(A_1|A_1)\right]A_{1}$. Hence $A_2(t)$ is orthogonal to both $A_1$ and $A_0$.
The Laplace transformation to Eqs.(52) and (53) reads
\begin{eqnarray}    \label{eq:54}
   z A_1(z) = A_1(0) + i \left[ Q_1 L A_1\right](z),
\end{eqnarray}
and
\begin{eqnarray}    \label{eq:55}
&& A_1(z)= \frac{A_1 + i A_2(z)}{z + i\Omega_1 + K_2(z)},
\end{eqnarray}
respectively.
Projecting Eq.(55) to $A_1$ gives
\begin{equation}  \label{eq:56}
   K_1(z) = \frac{(A_1\vert A_1)}{z+i\Omega_1 + K_2(z)}.
\end{equation}
Repeating the above derivation, we will obtain the CF for $K_0(z)$, as did by Mori\cite{RefMZ:19}, Lee\cite{RefLee}, Tserkovnikov\cite{RefTserkovnikov}, etc.

The specialty of the AIM is that some components in $A_{i}(z)$ ($i=1,2,...$), such as the bath operators $c_{k\sigma}(z)$, can be solved exactly from their Heisenberg EOM. A $z$-dependent term can thus be separated from the memory function, leading to an extended CF for $K_0(z)$. To employ this properties of the impurity model, we split the operator $A_1(z)$ as $A_1(z) = B_1(z) + C_1(z)$.  Both $B_1(z)$ and $C_1(z)$ have their own equation of motion and the generalized Langevin equation, similar to those of $A_1(z)$ in Eqs.(54) and (55). As to be shown below, from $A_1(z)$ we can collect the bath operators $c_{k\sigma}(z)$ into $B_1(z)$ and solve its EOM exactly. It provides the frequency-dependent term in the extended CF.   We need only to apply the generalized Langevin equation for $C_1(z)$ alone.

To guarantee the orthogonality between different hierarchy of operators, we also need $(A_0 \vert B_1)=(A_0 \vert C_1)=0$ to be fulfilled. The time-evolution is given by $B_1(t)=e^{iQ_1 Lt}B_1$ and $C_1(t)=e^{iQ_1 Lt}C_1$, similar to that of $A_1(t)$. We then have
\begin{eqnarray}   \label{eq:57}
K_0(z) = \frac{(A_0\vert A_0)}{z + i\Omega_0 + K_{1B}(z)+ K_{1C}(z)},
\end{eqnarray}
with
\begin{eqnarray}    \label{eq:58}
&& K_{1B}(z)=\frac{1}{(A_0\vert A_0)} \left[ (B_1\vert B_1(z))+(B_1\vert C_1(z))+(C_1\vert B_1(z)) \right],  \nonumber\\
&& K_{1C}(z)=\frac{1}{(A_0\vert A_0)} \left( C_1\vert C_1(z) \right).
\end{eqnarray}
Once $B_1(z)$ is solved exactly from its equation of motion $zB_1(z) = B_1 + i \left[Q_1 L B_1 \right](z)$, it is easy to calculate $(B_1\vert B_1(z))$ and $(C_1\vert B_1(z))$. $(B_1\vert C_1(z))$ can be calculated from $(B_1\vert C_1(z))=\int_{0}^{\infty} dt e^{-zt}(B_1(-t)\vert C_1)$. So $K_{1B}(z)$ is easily calculated. 
For $K_{1C}(z)$, we write down the generalized Langevin equation for $C_1(z)$ as
\begin{equation}  \label{eq:59}
  C_1(z) = \frac{C_1 + i A_2(z)}{z + i \Omega_1 + K_2(z)},
\end{equation}
which gives 
\begin{equation}  \label{eq:60}
   K_{1C}(z) = \frac{1}{(A_0\vert A_0)} \frac{(C_1\vert C_1)}{z + i \Omega_1 + K_2(z)}.
\end{equation}
Here, the frequency and memory functions are given by
\begin{eqnarray}  \label{eq:61}
   \Omega_1 = \frac{(C_1| \left[C_1, H\right])}{(C_1|C_1)} ,  \nonumber \\
   K_2(z) = \frac{(A_2|A_2(z))}{(C_1|C_1)}.
\end{eqnarray}
The second-order memory function $K_2(z)$ is defined by the random force $A_2(z)$ of $C_1(z)$. $A_2(z)$ is the Laplace transformation of $A_2(t)=e^{iQ_2Q_1 Lt}A_2$, with $A_2 = Q_2 Q_1L C_1$. Here $Q_2=1-P_2$ and $P_2 O = [(C_1|O)/(C_1|C_1)] C_1$. Therefore, $A_2(z)$ is orthogonal to $C_1$. 

The same procedure can be implemented for $A_2(z)$, i.e., from $A_2(z)$ we collect the bath operators into $B_2(z)$ and do the splitting $A_2(z) = B_2(z) + C_2(z)$. We apply the equation of motion and generalized Langevin equation to $B_2(z)$ and $C_2(z)$, respectively. Employing the exactly solved $B_2(z)$, we obtain the second-order memory function $K_2(z)= K_{2B}(z)+K_{2C}(z)$. $K_{2B}(z)$ is an exactly solvable frequency-dependent term
\begin{eqnarray}   \label{eq:62}
&& K_{2B}(z)= \frac{1}{(C_1\vert C_1)}[(B_2\vert B_2(z))+(B_2\vert C_2(z))+(C_2\vert B_2(z))] , \nonumber\\
\end{eqnarray}
and $K_{2C}(z)$ is an unknown memory function
\begin{eqnarray}  \label{eq:63}
&& K_{2C}(z)= \frac{1}{(C_1\vert C_1)}(C_2\vert C_2(z)).
\end{eqnarray}
Here, we also require that $(C_1\vert B_2)=(C_1\vert C_2)=0$. The time evolution is given by $B_2(t)=e^{iQ_2Q_1 Lt}B_2$ and  $C_2(t)=e^{iQ_2Q_1 Lt}C_2$.
By repeating the above process, we can express $K_0(z)$ into an infinite extended CF.

For the Hamiltonian in Eq.(\ref{eq:12}), we start from $A_0 = d_{\sigma}$ and obtain the following operators through straightforward calculation,
\begin{eqnarray}   \label{eq:64}
&& B_1= -\sum_k V_{k\sigma}c_{k\sigma},   \nonumber  \\ 
&& C_1= -U[n_{\bar{\sigma}}-\langle n_{\bar{\sigma}}\rangle]d_{\sigma},  \nonumber\\
&& B_2= -U\langle n_{\bar{\sigma}}\rangle \sum_k V_{k\sigma}c_{k\sigma}, \nonumber \\
&& C_2= U\sum_k V_{k\sigma}n_{\bar{\sigma}}c_{k\sigma}+U\sum_k V_{k\bar{\sigma}}(d_{\bar{\sigma}}^{\dagger}c_{k\bar{\sigma}}d_{\sigma}-c_{k\bar{\sigma}}^{\dagger}d_{\bar{\sigma}}d_{\sigma}) \nonumber\\
&& \hspace{2em}- U \tilde{\beta}_{\sigma} \left[ n_{\bar{\sigma}}-\langle n_{\bar{\sigma}}\rangle \right]d_{\sigma}. 
\end{eqnarray}
Here $\tilde{\beta}_{\sigma}$ is given in Eq.(35). They fulfil the orthogonal requirements and the EOM of $B_1(z)$ and $B_2(z)$ can be solved exactly. From these operators, we obtain the projecting coefficients used in the extended CF
\begin{eqnarray}    \label{eq:65}
&& \Omega_{0}= \epsilon_d + U\langle n_{\bar{\sigma}}\rangle,  \nonumber \\
&&\Omega_{1}=\epsilon_d + U(1-\langle n_{\bar{\sigma}}\rangle) + \tilde{\beta}_{\sigma},  \nonumber\\
&&(C_1\vert C_1)= U^2 \langle n_{\bar{\sigma}} \rangle(1-\langle n_{\bar{\sigma}}\rangle), \nonumber\\
&& K_{1B}(z)= \sum_k\frac{V_{k\sigma}^2}{z+ i\epsilon_{k\sigma} },  \nonumber \\
&& K_{2B}(z)= -\frac{\langle n_{\bar{\sigma}}\rangle}{1-\langle n_{\bar{\sigma}}\rangle} \sum_k\frac{V_{k\sigma}^2}{z+i \epsilon_{k\sigma}},   \nonumber\\
&& K_{2C}(z) =\frac{(C_2\vert C_2(z))}{U^2\langle n_{\bar{\sigma}}\rangle(1-\langle n_{\bar{\sigma}}\rangle)}. 
\end{eqnarray}
Finally, putting Eq.(65) into Eqs.(57), (58), (60), (62), and (63), we obtain the two-level extended CF Eq.(34).


\begin{thebibliography}{}
\bibitem{RefPW:1} P. W. Anderson, Phys. Rev. \textbf{124}, 41 (1961). 

\bibitem{RefAC:2} A. C. Hewson, \textit{The Kondo Problem to Heavy Fermions} (Cambridge University Press, Cambridge, England, 1993).

\bibitem{RefTK:3} T. K. Ng and P. A. Lee, Phys. Rev. Lett. \textbf{61}, 1768 (1988).

\bibitem{RefLI:4} L. I. Glazman and M. E. Raikh, JETP Lett. \textbf{47}, 452 (1988).

\bibitem{RefYM:4} Y. Meir, N. S. Wingreen, and P. A. Lee, Phys. Rev. Lett. \textbf{70}, 2601 (1993). 

\bibitem{RefRZ:5} R. \v{Z}itko, and J. Bon\v{c}a, Phys. Rev. B \textbf{74},  045312 (2006).

\bibitem{RefRH:5} R. Hanson, L. P. Kouwenhoven, J. R. Petta, S. Tarucha, and L. M. K.Vandersypen, Rev. Mod. Phys. \textbf{79}, 1217 (2007).

\bibitem{RefRB:6} R. Brako, and D. M. Newns, J. Phys. C: Solid State Phys. \textbf{14}, 3065 (1981).

\bibitem{RefDC:7} D. C. Langreth, and P. Nordlander, Phys. Rev. B \textbf{43}, 2541 (1991).

\bibitem{RefAG:8} A. Georges, G. Kotliar, W. Krauth, and M. J. Rozenberg, Rev. Mod. Phys. \textbf{68}, 13 (1996).

\bibitem{RefPB:6} P. B. Wiegmann and A. M. Tsvelick,  J. Phys. C: Solid State Phys. \textbf{16}, 2281 (1983).

\bibitem{RefKY:9} K. Yosida and K. Yamada, Prog. Theor. Phys. \textbf{46}, 244 (1970); K. Yamada, Prog. Theor. Phys. \textbf{53}, 970 (1975).

\bibitem{RefSE:10} S. E. Barnes, J. Phys. F: Met. Phys. \textbf{6}, 1375 (1976).

\bibitem{RefCL:11} C. Lacroix, J. Phys. F: Met. Phys. \textbf{11}, 2389 (1981). 

\bibitem{RefHG:12} H. G. Luo, J. J. Ying, and S. J. Wang, Phys. Rev. B \textbf{59}, 9710 (1999).

\bibitem{RefJK:13} J. Kroha, P. W\"{o}lfle, Advances in Solid State Physics \textbf{39}, 271 (1999).

\bibitem{RefNE:13} N. E. Bickers, Rev. Mod. Phys. \textbf{59}, 845 (1987).

\bibitem{RefKH:14} K. Haule, S. Kirchner, J. Kroha, and P. W\"{o}lfle, Phys. Rev. B \textbf{64}, 155111 (2001).

\bibitem{RefJE:15} J. E. Hirsch and R. M. Fye, Phys. Rev. Lett. \textbf{56}, 2521 (1986);  R. M. Fye and J. E. Hirsch, Phys. Rev. B \textbf{38}, 433 (1988).

\bibitem{RefMF:16} M. Feldbacher, K. Held, and F. F. Assaad, Phys. Rev. Lett. \textbf{93}, 136405 (2004).

\bibitem{RefEG:17} E. Gull, A. J. Millis, A. I. Lichtenstein, A. N. Rubtsov, M. Troyer, and P. Werner, Rev. Mod. Phys. \textbf{83}, 349 (2011).

\bibitem{RefHR:18} H. R. Krishna-murthy, J. W. Wilkins, and K. G. Wilson, Phys. Rev. B \textbf{21}, 1003 (1980); Phys. Rev. B 
\textbf{21}, 1044 (1980).

\bibitem{RefTA:19} T. A. Costi, A. C. Hewson, and V. Zlatic, J. Phys.: Condens. Matter. \textbf{6}, 2519 (1994). 

\bibitem{RefRB:20} R. Bulla, A. C. Hewson, and T. Pruschke, J. Phys.: Condens. Matter. \textbf{10}, 8365 (1998).

\bibitem{RefRB:21} R. Bulla, T. A. Costi, and T. Pruschke, Rev. Mod. Phys. \textbf{80}, 395 (2008).

\bibitem{RefZH:22} Z. H. Li, N. H. Tong, J. H. Wei, D. Hou, X. Zheng, J. Hu, and Y. J. Yan, Phys. Rev. Lett. \textbf{109}, 266403 (2012).

\bibitem{RefNH:22} N. H. Tong, Phys. Rev. B \textbf{92}, 165126 (2015).

\bibitem{RefMZ:19} H. Mori, Prog. Theor. Phys.\textbf{33}, 423 (1965); \textbf{34}, 399 (1965); S. Nordholm and R. Zwanzig, J. Stat. Phys. \textbf{13}, 347 (1975).

\bibitem{RefSP:23} S. Pairault, D. S\'{e}n\'{e}chal, and A.-M. S. Tremblay, Phys. Rev. Lett. \textbf{80}, 5389 (1998); S. Pairault, D. S\'{e}n\'{e}chal, and A.-M. S. Tremblay, Eur. Phys. J. B \textbf{16}, 85 (2000).

\bibitem{RefJG:24} J. Gilewicz, \textit{Approximants de Pad\'{e}},  Lecture Notes in Mathematics \textbf{667} (Springer-Verlag, Berlin, 1978.)

\bibitem{RefSB:25} S. B. Haley, Phys. Rev. B \textbf{17}, 337 (1978).

\bibitem{RefHL:26} H. Li and N. H. Tong, Eur. Phys. J. B \textbf{88}, 319 (2015).

\bibitem{Fan1} P. Fan, K. Yang, K. H. Ma, and N. H. Tong, Phys. Rev. B {\bf 97}, 165140 (2018).

\bibitem{Fan2} P. Fan and N. H. Tong, arXiv:1812.05906.

\bibitem{RefAW:27} A. Weichselbaum and J. von Delft, Phys. Rev. Lett. \textbf{99}, 076402 (2007); R. Peters, T. Pruschke, and F. B. Anders, Phys. Rev. B \textbf{74}, 245114 (2006).

\bibitem{RefZH} The accuracy of NRG results for AIM was discussed in the Supplemental Material of Ref.\cite{RefZH:22}.

\bibitem{RefTP:27} T. Pruschke, M. Jarrell, and J. K. Freericks, Adv. Phys. \textbf{44}, 187 (1995); R. Zitzler, T. Pruschke, and R. Bulla,  Eur. Phys. J. B \textbf{27}, 473 (2002); R. Zitzler, N. H. Tong, T. Pruschke, and R. Bulla, Phys. Rev. Lett. {\bf 93}, 016406 (2004); R. Zitzler, {\it Magnetic Properties of the One-Band Hubbard Model}, Ph.D thesis (2004).

\bibitem{RefGebhard} F. Gebhard, {\it The Mott Metal-Insulator Transition} (Springer -Verlag Berlin Heidelberg New York, 1997.)

\bibitem{RefQMC:28} M. Jarrell, Phys. Rev. Lett. \textbf{69}, 168 (1992); A. Georges and W. Krauth, Phys. Rev. B \textbf{48}, 7167 (1993); M. J. Rozenberg, G. Kotliar, and X. Y. Zhang, Phys. Rev. B \textbf{49}, 10181 (1994).

\bibitem{RefLee} M. H. Lee, Phys. Rev. Lett. {\bf 49}, 1072 (1982); Phys. Rev. B {\bf 26}, 2547 (1982).

\bibitem{RefHong} J. Hong and H. Y. Kee, Phys. Rev. B {\bf 52}, 2415 (1992); {\bf 62}, 12581 (2000).

\bibitem{RefTserkovnikov} Yu. A. Tserkovnikov, Theor. Math. Phys. {\bf 49}, 993 (1981); {\bf 50}, 171 (1982); {\bf 118}, 85 (1999).  

\end{thebibliography}
\end{document}